\documentclass[fleqn,usenatbib]{mnras}

\usepackage{newtxtext,newtxmath}

\usepackage[T1]{fontenc}

\DeclareRobustCommand{\VAN}[3]{#2}
\let\VANthebibliography\thebibliography
\def\thebibliography{\DeclareRobustCommand{\VAN}[3]{##3}\VANthebibliography}

\usepackage{graphicx}	
\usepackage{amsmath}	
\usepackage{rotating}

\title[Shock-accelerated outflows in NGC 1068]{Dust destruction signals shock-accelerated outflows in the nearby active galaxy NGC\;1068}

\author[Luke R. Holden et. al]
{Luke R. Holden$^{1}$,\thanks{E-mail: l.holden@herts.ac.uk}
Clive Tadhunter$^{2}$,
Daniel J. B. Smith$^{1}$,
Martin A. Bourne$^{1,3}$, \newauthor
Marina I. Arnaudova$^{4}$,
Isaac M. Mutie$^{5,6}$
\\
$^{1}$Centre for Astrophysics Research, University of Hertfordshire, Hatfield, AL10 9AB, UK. \\
$^{2}$School of Mathematics and Physical Sciences, University of Sheffield, S6 3TG Sheffield, UK. \\
$^{3}$Kavli Institute for Cosmology, University of Cambridge, Cambridge, CB3 0HA, UK. \\
$^{4}$Institute for Astronomy, University of Edinburgh, Edinburgh, EH9 3HJ, UK. \\
$^{5}$Department of Astronomy and Space Science, Technical University of Kenya, Nairobi, PO Box 52428-00200, Kenya. \\
$^{6}$Jodrell Bank Centre for Astrophysics, University of Manchester, Manchester, M13 9PL, UK. \\
}

\date{Accepted XXX. Received YYY; in original form ZZZ}

\pubyear{2025}

\begin{document}
\label{firstpage}
\pagerange{\pageref{firstpage}--\pageref{lastpage}}
\maketitle

\begin{abstract}
Massive gas outflows driven by active galactic nuclei (AGN) are a key ingredient in models of galaxy evolution, in which they are required to regulate star formation and thus explain the observed properties of the galaxy population. However, it remains uncertain how such outflows are accelerated. Here, we use deep spectroscopic observations of the nearby active galaxy NGC\,1068 to directly address this issue. Based on the flux ratios of high-ionisation [NeV]$\lambda$3425 and [FeVII]$\lambda$6087 coronal forbidden lines, we show that the non-outflowing gas in the disk of the galaxy is characterised by high levels of depletion of refractory elements onto dust grains, but the outflowing gas just above the disk is largely dust-free. Consistent results are also found for the ratios of lower-ionisation forbidden lines of refractory and non-refractory elements.  Moreover, a range of diagnostic ratios demonstrate that the density of outflowing gas is a factor 19--110 times higher than that of the non-outflowing gas. Together, these results imply that the outflows in NGC\,1068 are accelerated by fast shocks that both compress the gas and destroy much of the dust. Consistent with the idea that AGN-driven shocks play an important role in heating and accelerating the near-nuclear gas in galaxies, this study demonstrates that coronal emission lines are a key diagnostic of the destructive impact of AGN activity.
\end{abstract}




\section{Introduction}
\label{section: introduction}

Outflows  of warm ($T\sim10^4$\;K) gas are often observed on 10\,pc -- 5\,kpc scales in the narrow-line regions (NLRs) of AGN, where they are detected as high-velocity wings ($\Delta V\sim$300 -- 2000\,km s$^{-1}$) in the profiles of forbidden emission lines \citep{Heckman1981, Whittle1988, Mullaney2013}. Such outflows are often massive, and in some cases sufficiently energetic to affect the evolution of their host galaxies \citep{Fiore2017, Rose2018, Santoro2018, Harrison2018, Fluetsch2019}. Nevertheless, for a complete picture, it is essential to understand how the outflows are accelerated. Without such understanding, theoretical models of galaxy evolution cannot implement accurate feedback prescriptions and are therefore unable to make robust, testable predictions.

Both direct radiation pressure and shocks driven by jets or accretion-disk winds have been suggested as possible driving mechanisms. While evidence for the radiation pressure mechanism is provided by detailed modelling of the spatially-resolved emission-line kinematics in the NLRs of nearby Seyfert galaxies \citep{Das2006,Fischer2017,Meena2023}, shock-acceleration is supported by several lines of evidence. These include the spatial coincidence of regions of kinematic disturbance with synchrotron-emitting radio jets and lobes in some objects \citep{Whittle1988,Axon1998,Capetti1999,Morganti2007, Audibert2023, Holden2024}, correlations between emission-line widths and radio luminosities \citep{Whittle1992c, Mullaney2013}, and observations of fast (0.1--0.2\;$c$), tens-of-parsec scale winds in X-ray emission \citep{Tombesi2011, Feruglio2015, KingPounds2015}. However, such results are often indirect or model-dependent, and despite a wealth of literature on warm outflows in the NLRs of AGN, it has proved challenging to distinguish between the mechanisms definitively \citep[e.g.][]{Rosario2010a, HoldenTadhunter2023}. 

Much of the uncertainty regarding acceleration mechanisms persists because previous work did not isolate the properties of the outflows from those of the non-outflowing gas, preventing key information about the nature of the acceleration from being inferred. To directly address this, here we present an analysis of deep Very Large Telescope (VLT)/Xshooter long-slit spectroscopy of the nearby Seyfert galaxy NGC\,1068. Concentrating on a single region 230\,pc to the north-east of the nucleus in the NLR of NGC\,1068, we investigate the spectroscopic evidence for dust destruction and gas compression, both of which provide key information about the outflow acceleration mechanism.

Due to its proximity ($d=13$\,Mpc: \citealt{Schoniger94}) and status as the prototypical Type 2 Seyfert galaxy, NGC\;1068 has been extensively studied across the electromagnetic spectrum on a range of spatial scales \citep[e.g.][]{Seyfert1943,Wilson1983,Young2001,Honig2008,GarciaBurillo2014, Gravity2020, Revalski2021, Marconcini2025}. In addition to a relatively high luminosity AGN ($L_{bol} = (0.4$--$4.7)\times10^{45}$\,erg\;s$^{-1}$: \citealt{Gravity2020}), it hosts a kiloparsec-scale radio source of moderate luminosity ($L_\mathrm{1.4\;GHz}=1.2\times10^{23}$\,W Hz$^{-1}$: \citealt{Ulvestad1984}) that has a similar radial extent to the bulk of the multi-phase outflows that are detected in its NLR \citep{HoldenTadhunter2023}. For these reasons, it is a key object for understanding AGN outflows and how they are accelerated.

The paper is organised as follows:  in Section \ref{section: observations_and_data_preparation} we describe the Xshooter observations and data reduction process; Section\;\ref{section: spectral_fitting} details our aperture selection and spectral-fitting methodology, and Section\;\ref{section: results} presents the results of our analysis using various emission-line diagnostics. Finally, we discuss these results in Section\;\ref{section: discussion}, and give our conclusions in Section\;\ref{section: conclusions}. Throughout, we assume a distance to NGC\;1068 of 13.0\,Mpc, based on the CO and HI Tully-Fisher estimates of \citet{Schoniger94}; this corresponds to a spatial scale of 63\,pc\;arcsec$^{-1}$. 

\section{Observations and data reduction}
\label{section: observations_and_data_preparation}

\begin{figure*}
	\includegraphics[width=\linewidth]{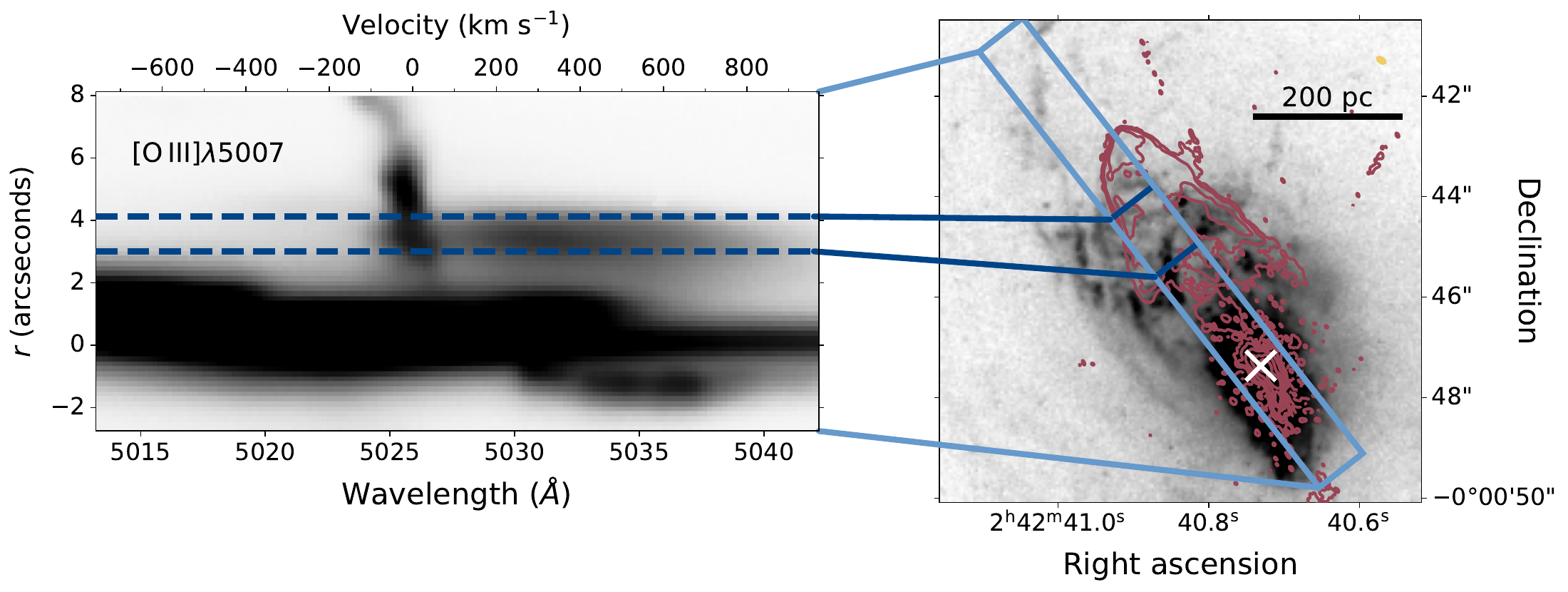}
	\caption{Left: the two-dimensional spectrum of NGC\;1068 in the wavelength region of the $\mathrm{[O\,III]}\lambda5007$ emission line (see Figure\;\ref{fig: oiii_xshooter_full} for the entire $\mathrm{[O\,III]}\lambda\lambda4959,5007$ doublet profile), as seen along the slit position marked in light blue on the right panel; the dashed dark-blue lines show the spatial extent of the region at the base of the radio lobe that we extracted from this spectrum. Right: the nuclear region of NGC\;1068 as seen in Hubble Space Telescope [O\,III] narrow-band imaging (black with a white background) and contours of a radio map derived from a combination of 5, 10, and 21\;GHz data (red); the white cross shows the position that we take to be that of the nucleus, and the yellow ellipse in the top right indicates the beam size of the combined radio image. Lupton flux scaling \citep{Lupton2004} is used in both panels for presentation purposes.}
	\label{fig: slit_position}
\end{figure*}

NGC\;1068 was observed on the 24th of August 2023 with the Xshooter long-slit spectrograph mounted on Unit Telescope 2 of the VLT as part of program ID 111.24FW.001 (PI Holden)\footnote{Our VLT/Xshooter data are available from the ESO Science Archive Facility: \url{https://archive.eso.org/eso/eso_archive_main.html}}, with separate sky exposures taken in a nodding object-sky-sky-object observing pattern. The 1.0\;arcsecond-width slit was aligned along position angle ($\mathrm{PA})=38^\circ$ --- corresponding to the axis of the radio structure of NGC\;1068 --- with the slit centred on a bright, compact emission-line region close to the expected position of the object's active nucleus (the `S1' radio component as labelled by \citealt{Gallimore1996}; see also  \citealt{Kraemer2000II}). Moreover, we ensured that the slit length contained the full extent ($r\approx6.2\mathrm{\;arcseconds}\approx390$\;pc) of the radio lobe in the northeastern (NE) cone of the narrow line region (see Figure\;\ref{fig: slit_position}); the full slit covered 2.6\;arcseconds south of the nucleus and 8.3\;arcseconds to the north. The total on-object exposure time was 800\;s (consisting of $8\times100$\;s exposures) for the UVB (3000-5600\;{\AA}; $R=5400$; $0.161$\;arcseconds per pixel) and VIS (5500-10200\;{\AA}; $R=8900$; $0.158$\;arcseconds per pixel) arms of the spectrograph. The average atmospheric seeing during the observations was 0.58\;arcseconds (FWHM), as measured with the observatory's Differential Image Motion Monitor.

The Xshooter UVB and VIS data were reduced using the Xshooter pipeline (version 3.6.7) with version 2.11.5 of \textsc{EsoReflex} \citep{Freudling2013}, which processed the data and performed bias correction, dark subtraction, flat fielding, flux and wavelength calibration, and combined the constituent exposures into a single two-dimensional spectrum for each arm. To interpolate over bad pixels in the resulting spectra (for example, those affected by cosmic ray hits), we ran two passes of a median filter of size 5\;px using the \textsc{SciPy Python} package \citep{Virtanen2020}. Correction of telluric absorption lines was performed using \textsc{molecfit}. Finally, the UVB and VIS spectra were corrected for Galactic extinction using the \citet{Cardelli1989} extinction law and an assumed Galactic reddening in the direction of NGC\,1068 of $\mathrm{E(B-V)}=0.0289$, as derived from the maps of \citet{Schlafly2011}.

After taking into account a 1.4 pixel shift in the spatial direction of the reduced long-slit spectra between the UVB and VIS arms, the flux-density values for the constituent pixel rows in our chosen region were summed, and the flux errors from the Xshooter pipeline added in quadrature. In order to check that the flux calibration was consistent between the arms, we measured the median flux measured by each in the overlap wavelength region (5500--5520\;{\AA}). We found that the UVB flux in this region was 5\;per\;cent higher, and therefore corrected this by multiplying all the flux values for the VIS arm by a factor of 1.05. 

\begin{figure*}
	\centering
	\includegraphics[width=\linewidth]{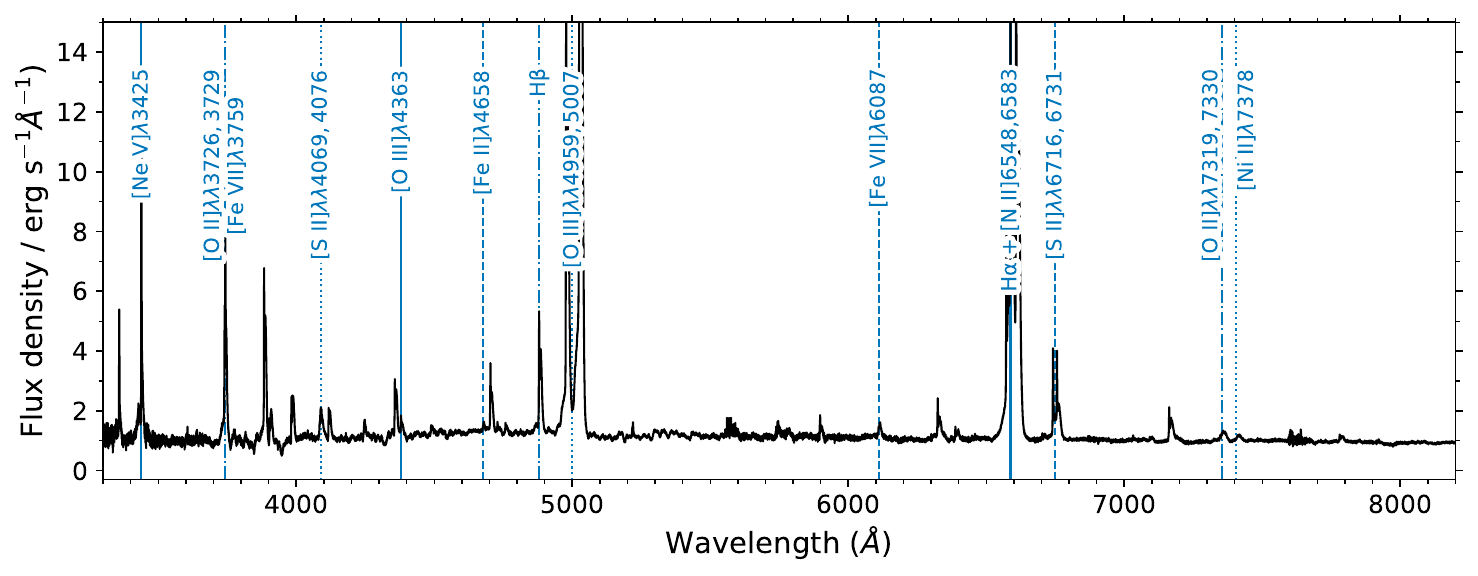}
	\caption{The spectrum extracted from the UVB and VIS Xshooter arms at the position of the base of northeastern radio lobe in NGC\;1068 (see Figure\;\ref{fig: slit_position}), corresponding to our spectroscopic aperture. For presentation purposes, the wavelength range here has been chosen to cover all key diagnostic emission lines that we use in our study (labelled), while the flux-axis limits were chosen to show the fainter lines (bright lines such as $\mathrm{[O\,III]}\lambda\lambda4959,5007$ extend beyond these limits).}
	\label{fig: radio_lobe_spectrum_full}
\end{figure*}

We measured the instrumental full-width at half maximum (FWHM) in the spectral direction using Galactic ISM absorption lines detected in our spectra: the $\mathrm{Ca\,II\,K}\lambda3934$ absorption feature in the UVB arm, and the $\mathrm{Na\,I\,D}\lambda\lambda5890,5896$ doublet in the VIS arm. This was done by fitting these lines with single Gaussian profiles for spectra extracted from a region south of the nucleus that was free of significant line emission. In this way, we determined instrumental-broadening FWHMs of $0.70\pm0.05$\;{\AA} and $0.72\pm0.02$\;{\AA} for the UVB and VIS arms, respectively.

The radio-continuum image of the near-nuclear region of NGC\;1068 --- shown as red contours superimposed on the grey-scale [O\,III] optical image and our Xshooter slit position in Figure\;\ref{fig: slit_position} --- was obtained by combining the  e-MERLIN (enhanced Multi Element Remotely Linked Interferometer Network) and VLA (Karl G. Jansky Very Large Array) observations at 5, 10, and 21\;GHz of the central region of NGC\;1068 that were originally presented by \citet{Mutie2024}, to which we direct readers for a full description of the observations and data reduction. The narrow-band [O\,III] image that is presented as the black-and-white background image in Figure\;\ref{fig: slit_position} was taken with the F502N filter of the Hubble Space Telescope/Wide Field Planetary Camera 2 (HST/WFPC2) instrument (proposal ID 5754, PI Ford)\footnote{The HST/WFPC2 data used here are available from the Hubble Legacy Archive: \url{https://hla.stsci.edu/}}.

\section{Aperture selection and  spectral fitting}
\label{section: spectral_fitting}

\subsection{Aperture selection}

Figure\;\ref{fig: slit_position} shows the position of the Xshooter slit relative to the radio source and the warm ionised gas in the NLR, as well as a section of the long-slit spectrum in the vicinity of the $\mathrm{[O\,III]}\lambda5007$ emission line\footnote{See Figure \ref{fig: oiii_xshooter_full} in the Appendix for the full velocity range encompassed by the [OIII] lines.}. Since we are interested in comparing the properties of non-outflowing and outflowing gas at the same projected radial distance from the central AGN, we base our main analysis on a spectrum extracted from a 6\;pixel-wide (0.97\;arcseconds, 61\,pc) aperture centred 3.6\,arcsec ($230$\;pc) from the nucleus, corresponding to the base of the northeastern (NE) radio lobe. The extracted spectrum is presented in Figure\;\ref{fig: radio_lobe_spectrum_full}, with the displayed wavelength range chosen to cover the emission lines that we use in our analysis. These include a range of density-diagnostic lines as well as both low- ($\text{[Fe\;II]}\lambda4658$, $\text{[Ni\;II]}\lambda7378$) and high-ionisation (coronal: $E_\mathrm{ion}\gtrapprox100$\;eV; $\text{[Ne\;V]}\lambda3425$, $\text{[Fe\;VII]}\lambda3759$, $\text{[Fe\;VII]}\lambda6087$) lines, including those of refractory elements (those that most readily condense onto dust grains) such as iron and nickel.

\begin{figure}
    \vspace{1cm}    
	\includegraphics[width=\linewidth]{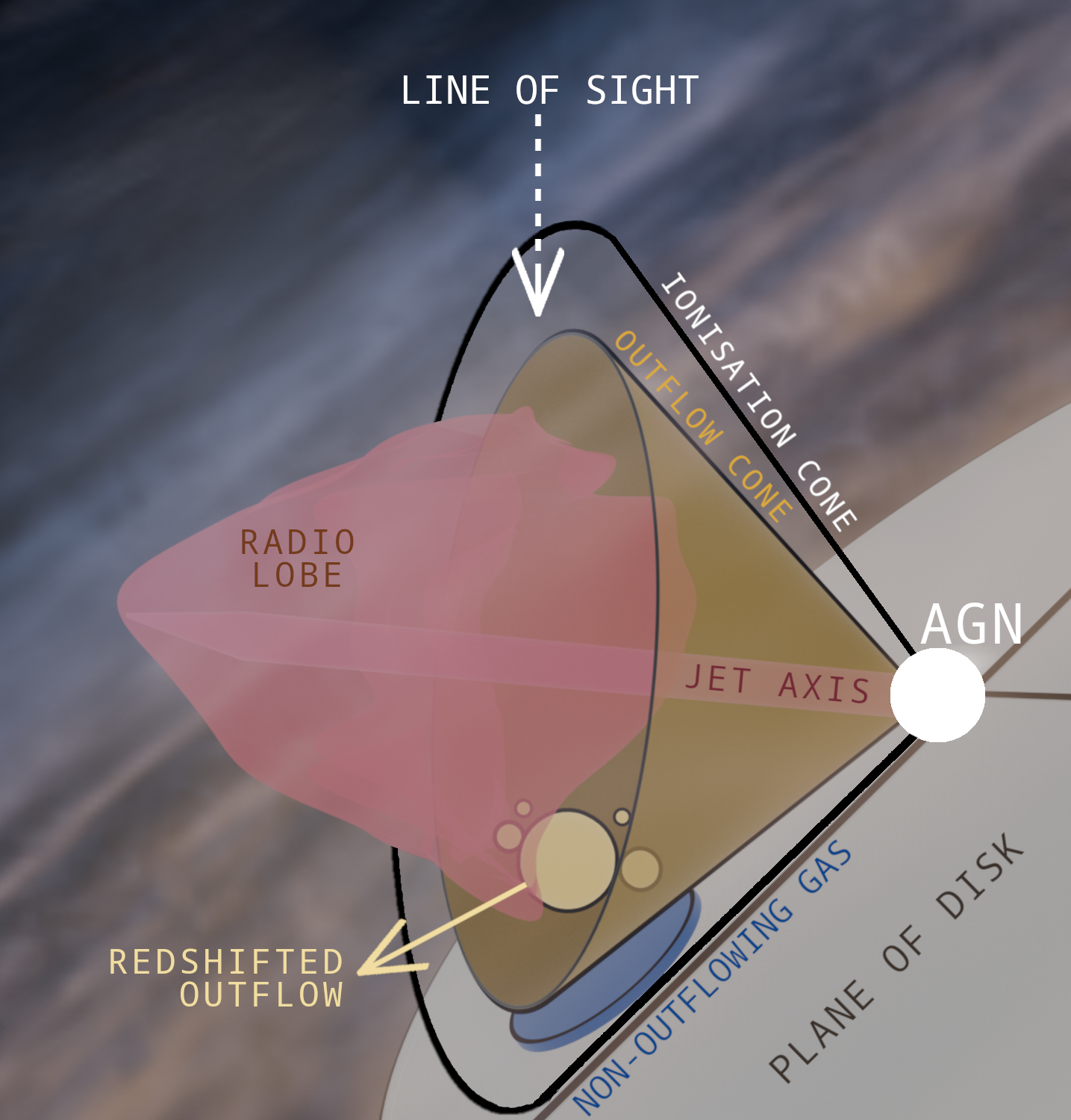}
	\caption{A representation of the kiloparsec-scale region to the north east of the nucleus of NGC\;1068. The background shows the inner disk and the illuminating AGN (with its ionisation cone shown as a solid black outline); the radio lobe is shown in red; a representation of the outflowing gas detected in our data is shown as yellow spheres (with the outflow cone from kinematic modelling by \citealt{Das2007} shown as a dark yellow cone), and the non-outflowing disk gas that is seen in our Xshooter observations is represented by the blue region. Note that while the positions and relative orientations of each element are informed by our observations and previous studies, they are not to scale. On the north-east side of the galaxy, the disk is known to be inclined from the plane of the sky by 40$\pm$3 degrees away from the observer \citep[e.g.][]{GarciaBurillo2014}, while according to the kinematic model of \citet{Das2006} the outflow cone has an opening half-angle of 40$\pm$2 degrees, and its axis is inclined from the plane of the sky in the direction of the observer by 5$\pm$2 degrees.}
    \vspace{1cm}
	\label{fig: ngc1068_model}
\end{figure}

The location of our spectroscopic aperture lies between the outflow-dominated inner part of the NLR ($r<150$\;pc) and the outer NLR ($r>300$\;pc), which is dominated by the emission of the kinematically-quiescent disk of the galaxy \citep{GarciaBurillo2014}. It was chosen because the narrow and broad emission-line profile components are clearly distinguishable: in addition to strong emission from a narrow, non-outflow component of width $\mathrm{FWHM} < 100$\;km\;s$^{-1}$, the line profiles show a broad, redshifted outflow component ($\Delta V\sim250$\,km s$^{-1}$; $\mathrm{FWHM}\sim600$\;km\;s$^{-1}$) that lacks the complexity of the profiles observed at smaller radii and is clearly distinct from the narrow component. According to the hollow bicone model for the gas kinematics in NGC\;1068 \citep{Das2006}, at this location we are observing a combination of emission from non-outflowing gas in the galaxy's disk (narrow component) and the outflowing gas (broad component) that lies just above it, at the edge of the far side of the NE outflow cone as seen from Earth. A schematic of this situation is shown in Figure\;\ref{fig: ngc1068_model}.


\subsection{Nebular and stellar continuum modelling and subtraction}
\label{section: spectral_fitting: continuum_modelling}

Prior to measuring the emission lines, we subtracted a nebular continuum from the UVB and VIS spectra using the output of the  \textsc{STARLINK DIPSO} \citep{Currie2014} \textsc{nebcont} routine, and then modeled the remaining starlight using the \textsc{pPXF} code \citep{Cappellari2004, Cappellari2017, Cappellari2023} with the Xshooter Spectral Library \citep{Verro2022a, Verro2022b} stellar templates after masking emission lines. \textsc{pPXF} was run with two groups of stellar components, each with distinct kinematics and multiple spectral templates. This was required to adequately describe the continuum in the intermediate-spectral-resolution Xshooter spectra. The fits to the Ca\,II K, Mg\,I, and G-band stellar absorption features were visually inspected to check overall quality of the fit, after which we subtracted the modeled stellar continuum from the aperture spectrum.

\subsection{Emission-line fitting}

We fitted the emission lines required for our analysis with the method and parameters described in \citet{HoldenTadhunter2025}, which involves Markov Chain Monte Carlo (MCMC) sampling with an iteratively increasing number of Gaussian components: the number of Gaussian components was increased until the change of the Bayesian Information Criterion (BIC) between iterations was less than six. Priors in our MCMC routine were set based on physical constraints --- the peaks and widths of each Gaussian component were required to be positive, and certain emission-line ratios were required to be in the range defined by atomic physics ($0.41<\mathrm{[O\;II](3729/3726)}<1.50$; $3.01<\mathrm{[S\;II](4068/4076)}<3.28$, $0.46<\mathrm{[S\;II](6717/6731)}<1.45$; determined using the \textsc{PyNeb Python} package: \citealt{Luridiana2015}). Furthermore, we fixed the flux ratios of the $\mathrm{[O\,II]}\lambda\lambda4959,5007$ and $\mathrm{[O\;II]}(7320/7331)$ doublets to be $1.00:2.99$ and $1.24:1.00$ respectively, the former based on the transition emissivities \citep{Osterbrock2006} and the latter based on previous photoionisation modelling \citep{Rose2018}. When fitting emission-line doublets, we assumed that two components of the doublet had the same velocity width, and a wavelength separation set by atomic physics.

\begin{figure*}
	\includegraphics[width=\linewidth]{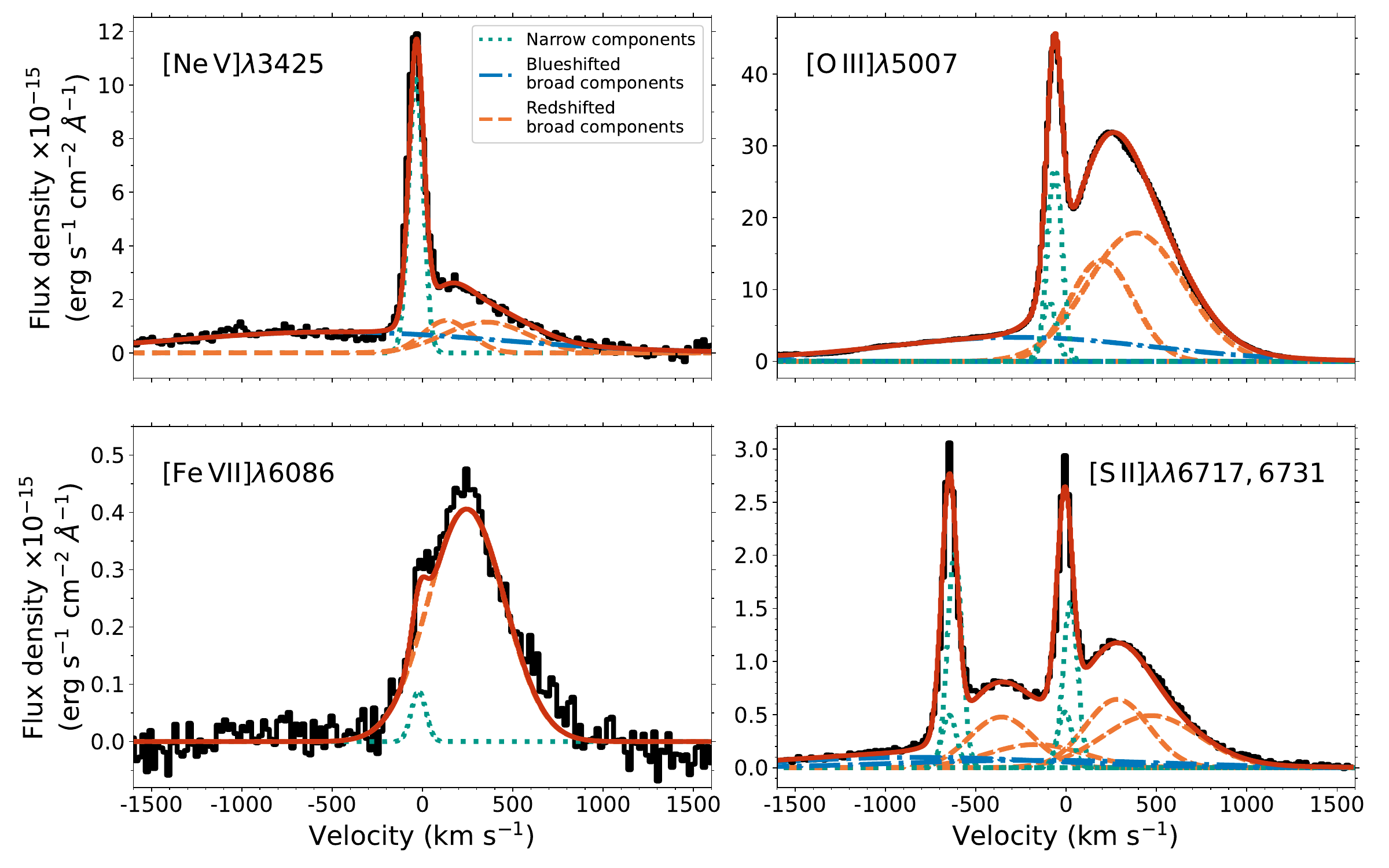}
	\caption{Spectral fits to key diagnostic lines that are used in our analysis (labelled in each panel). The extracted spectrum is shown as a solid black line; narrow components (which we take to represent non-outflowing gas: $\mathrm{FWHM}<200$\;km\;s$^{-1}$) are shown as dotted green lines; blueshifted broad components (an outflow: $\mathrm{FWHM}>200$\;km\;$^{-1}$) are shown as dash-dotted blue lines, and the redshifted broad components (the outflow that we consider in this work: $\mathrm{FWHM}>200$\;km\;s$^{-1}$; velocity $\mathrm{shift}>0$\;km\;s$^{-1}$) are shown as dashed orange lines. Here, the line profiles presented in the bottom panels have been mean-binned by a factor of two for presentation purposes.}
	\label{fig: spectral_fits}
\end{figure*}

We first fit the bright $\mathrm{[O\,III]}\lambda\lambda4959,5007$ doublet in order to constrain the kinematics of the fits to fainter lines: the model that was fit to this doublet was then fit to the other lines by fixing the velocity shifts and widths, but allowing the peak flux to vary, and adding and subtracting the instrumental width in quadrature as appropriate\footnote{All individual Gaussian components were broader than the instrumental width.}. We note that the broad components of the higher-ionisation [Ne\,V] and [Fe\,VII] emission-line profiles, although also redshifted, present different kinematics from the [O\,III] lines (indicating that they trace different parts of the same cloud complexes: \citealt{HoldenTadhunter2023}), and so we fixed only the narrowest component to have the same kinematics as the [O\,III] fits, while allowing independent Gaussian components to describe the broad components. The kinematics of the broad components of these lines, which are consistent with each other within $1\sigma$, are presented in Table\;\ref{tab: kinematics}. Since some of the fainter components of the $\mathrm{[O\,III]}\lambda\lambda4959,5007$ line-profile model were strongly affected by continuum noise in the case of other, fainter lines, we fit the Gaussian components iteratively in descending order of peak flux and evaluated the BIC upon adding each; following our earlier approach, additional components were only accepted if the change in BIC was greater than six. Example fits for several key diagnostic lines are shown in Figure\;\ref{fig: spectral_fits}. For all lines, we took the 50th percentile of the walker chains for each parameter as the parameter value, and half of the difference between the 16th and 84th percentiles as estimations of the $1\sigma$ uncertainties.

\begin{table}
	\centering
	\begin{tabular}{cccc}
		Line & $V_{50}$ (km\;s$^{-1}$) & $W_{80}$ (km\;s$^{-1}$) \\
		\hline 
		$\mathrm{[O\;III]}\lambda5007$ & $365\pm45$ & $756\pm94$ \\
		$\mathrm{[Ne\;V]}\lambda3425$ & $284\pm33$ & $578\pm72$ \\
		$\mathrm{[Fe\;VII]}\lambda3759$ & $238\pm17$ & $535\pm35$ \\
		$\mathrm{[Fe\;VII]}\lambda6086$ & $248\pm5$ & $542\pm35$ \\
            &   &   \\
	\end{tabular}
	\caption{Non-parametric measurements of the kinematics of the broad components of the $\mathrm{[O\;III]}\lambda5007$ emission-line-profile model and the high-ionisation coronal lines detected in the spectroscopic aperture considered in this work. The $V_{50}$ parameter is the velocity that contains 50\;per\;cent of the total emission-line flux (i.e. the velocity shift at the median of the line profile), while $W_{80}$ is the velocity width that contains 80\;per\;cent of the total flux. The kinematics for the coronal lines are consistent within $2\sigma$.}
	\label{tab: kinematics}
\end{table}

Based on previous kinematic modelling for the narrow line region (NLR) of NGC\;1068 (\citealt{Das2006}; see Figure\;\ref{fig: ngc1068_model}), we split the Gaussian components of the resulting emission-line fits for the $\mathrm{[O\,III]}\lambda\lambda4959,5007$ doublet into two categories depending on their kinematics: narrow components ($\mathrm{FWHM}<200$\;km\;s$^{-1}$) and high-velocity components ($\mathrm{FWHM}>200$\;km\;s$^{-1}$). The latter category was further separated into blueshifted and redshifted high-velocity components based on the measured velocity shifts (with the boundary at $v=0$\;km\;s$^{-1}$; see Figure\;\ref{fig: spectral_fits}). Considering the kinematical differences between these categories, we interpret the narrow category as being emitted by non-outflowing gas that is moving under regular rotational motions within the disk of the host spiral galaxy (hence the low velocity shifts and widths; see \citealt{GarciaBurillo2014}), and the two high-velocity categories as being emitted by distinct outflowing gas components that are moving towards and away from the observer, respectively. The blueshifted component detected in our extracted aperture is likely to represent seeing-disk spillover from the adjacent aperture closer to the nucleus (see Figure \ref{fig: slit_position}). Therefore, although we modelled this component to obtain the best possible fit to the overall line profile, we consider only the broad redshifted and the narrow components in our analysis.

\section{Results}
\label{section: results}

From the results of our spectral fits, it is clear that both the non-outflow and outflow kinematic components have a similarly high ionisation state, as indicated by emission-line flux ratios such as $\mathrm{He\,II}\lambda4686/\mathrm{H}\beta$, $\mathrm{[O\,III]}\lambda5007/\mathrm{H}\beta$, and $\mathrm{[O\,II]}\lambda3727/\mathrm{[O\,III]}\lambda5007$, which are presented in Table\;\ref{tab: line_ratios}. This is consistent with the results for the inner part of the NLR ($r<150$\;pc) that were derived from Hubble Space Telescope (HST) spectroscopy \citep{Crenshaw2000_N1068,HoldenTadhunter2023}, and supports AGN photoionisation as the main ionisation mechanism. Furthermore, the electron temperatures measured using the $\mathrm{[O\,III]}(4959+5007)/4363$ ratio are similar for the two components ($12400\pm200$\;K and $11100\pm200$\;K for the non-outflowing and outflowing gas, respectively). Given the consistency of these measured properties across several spectroscopic apertures, diagnostic ratios, kinematic components and studies, in the following we assume that AGN photoionisation is the dominant ionisation mechanism for the bulk of the warm, emission-line gas in the NLR of NGC\,1068.  However, despite the similarities in their ionisation properties, it is notable that the non-outflowing and outflowing gas differ in two important aspects: the electron density and level of depletion of refractory elements; both of these provide strong indications of the outflow acceleration mechanism.

\begin{table}
    \renewcommand{\arraystretch}{1.5}
    \centering
    \begin{tabular}{|c|c|c|}
    Ratio or parameter & Non-outflow & Outflow \\
    \hline
    $\mathrm{He\,II}(4686)/\mathrm{H}\beta$ & $0.60\pm0.08$ & $0.44\pm0.11$ \\
    $\mathrm{[O\,III]}(5007)/\mathrm{H}\beta$ & $10.9\pm2.6$ & $11.6\pm0.3$ \\
    $\mathrm{[O\,III]}(5007/4363)$ & $81.6\pm12.2$ & $106.0\pm5.3$ \\
    $\mathrm{[O\,II]}(3726)/\mathrm{[O\,III]}(5007)$ & $0.09\pm0.02$ & $0.05\pm0.01$ \\ 
    $\mathrm{[Fe\,VII]}(6086/(3759)$ & $3.18_{-0.77}^{+1.32}$ & $1.62_{-0.17}^{+0.17}$ \\
    $\mathrm{[Ne\,V]}\lambda3425/\mathrm{[Fe\,VII]}\lambda6086$ & $46.1^{+9.3}_{-9.5}$ & $3.18^{+0.02}_{-0.11}$ \\
    $\mathrm{[Ne\,V]}\lambda3425/\mathrm{[Ne\,III]}\lambda3869$ & $2.28\pm0.19$ & $0.47\pm0.04$ \\
    $\mathrm{[Fe\,II]}\lambda8617/\mathrm{[Ni\,II]}\lambda7378$ & --- & $0.84\pm0.21$ \\
    $\mathrm{[Si\,II]}(6717+6731)/\mathrm{[Ni\,II]}(7378)$ & --- & $6.94\pm1.75$ \\
    $\mathrm{[Ca\,II]}\lambda7291)/\mathrm{[O\,II]}(\lambda7319+\lambda7330)$ & --- & $<0.13$ \\
    $\mathrm{[Ca\,II]}\lambda7291/\mathrm{[Ni\,II]}\lambda7378$ & --- & $<0.22$ \\
    $\mathrm{H}\gamma/\mathrm{H}\beta$ & $0.42\pm0.04$ & $0.48\pm0.01$ \\
    $\mathrm{H}\delta/\mathrm{H}\beta$ & $0.22\pm0.02$ & $0.28\pm0.01$ \\
    $E(B-V)_{\rm TR}$ & $0.42^{+0.20}_{-0.19}$ & $0.00^{+0.04}_{-0.00}$ \\
    $E(B-V)_{\mathrm{H}\delta/\mathrm{H}\beta}$ & $0.20\pm0.02$ & $0.00\pm0.01$ \\
    $E(B-V)_{\mathrm{H}\gamma/\mathrm{H}\beta}$ & $0.22\pm0.02$ & $0.00\pm0.01$ \\
    $T_\mathrm{e}\;(\mathrm{K})$ & $12400\pm200$ & $11100\pm200$ \\
        &   &   \\
    \end{tabular}
    \caption{Diagnostic emission-line flux ratios and physical conditions for the narrow (non-outflowing) and broad (outflowing) gas components in the aperture that we extracted from our Xshooter spectrum of NGC\;1068. Electron temperatures were calculated using the [O\,III](4959+5007)/4363 ratio assuming electron densities that were derived from the transauroral-line technique (Section\;\ref{section: results: electron_density_measurements}; Figure\;\ref{fig: tr_grid}).}
    \label{tab: line_ratios}
\end{table}

\subsection{Electron density and reddening measurements}
\label{section: results: electron_density_measurements}

The depth of our Xshooter data allows us to measure the electron densities of the outflowing and non-outflowing gas in NGC\,1068 with precision. We first used the ``traditional'' $\mathrm{[S\,II]}(6717/6731)$ and $\mathrm{[O\,II]}(3729/3726)$ diagnostic flux ratios with the \textsc{PyNeb Python} package and an assumed electron temperature of $T_e = 10^4$\,K to measure the electron densities in our spectroscopic aperture. For the narrow component (corresponding to non-outflowing gas), we calculated densities of $n_{e, \mathrm{[S\;II]}}\;=288^{+92}_{-80}\,\mathrm{cm}^{-3}$ and 
$n_{e, \mathrm{[O\;II]}}\;=159^{+64}_{-27}\,\mathrm{cm}^{-3}$, whereas for the broad (outflowing) component, we measured a density of $n_{e, \mathrm{[S\;II]}}\;=5\,370^{+518}_{-472}\,\mathrm{cm}^{-3}$ (note that no [O\;II] density measurement was possible for the outflow component due to blending of the closely-spaced doublet lines). Therefore, using the $\mathrm{[S\,II]}(6717/6731)$ diagnostic we find that the outflow component has a density that is a factor $19^{+8}_{-5}$ higher than that of the non-outflow component; the contrast in the $\mathrm{[S\,II]}(6717/6731)$ ratio between the two kinematic components is clear in the doublet profile shown in the lower-right panel of Figure\;\ref{fig: spectral_fits}.

\begin{figure}
	\centering
	\includegraphics[width=\linewidth]{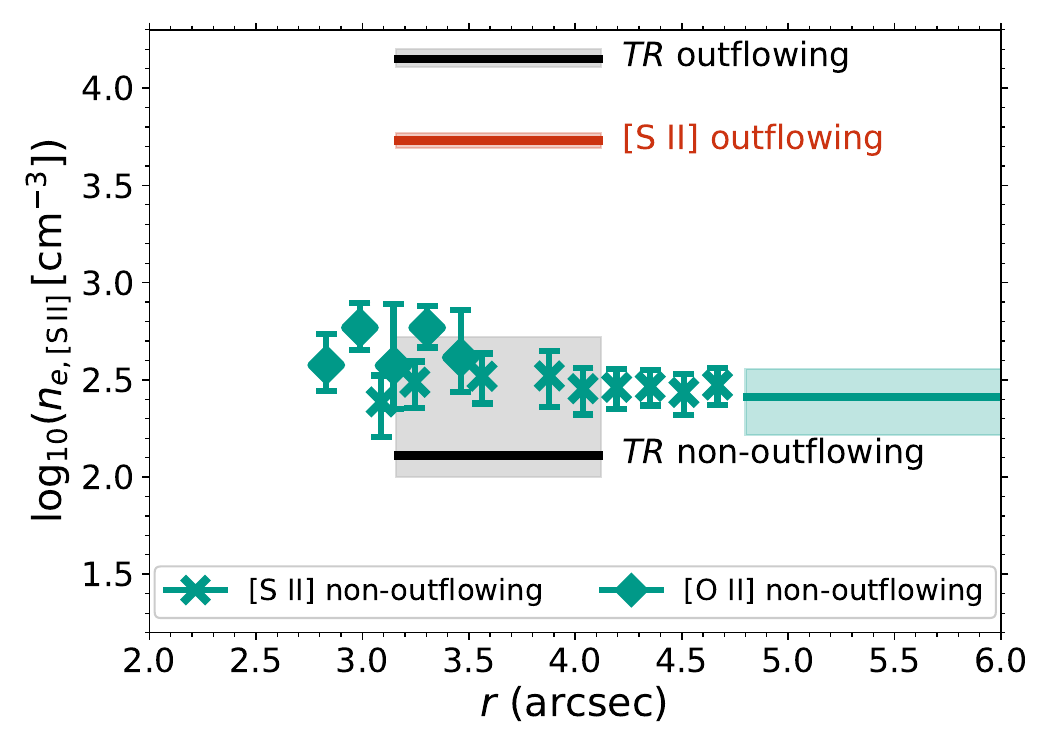}
	\caption{Electron-density values for the non-outflowing gas at increasing radial distance from the nucleus of NGC\;1068, as derived from the $\mathrm{[S\,II]}(6717/6731)$ and $\mathrm{[O\,II]}(3729/3726)$ flux ratios (coloured points; see legend); for comparison purposes, the outflowing and non-outflowing values measured for our extracted aperture (see Figure\;\ref{fig: slit_position}) using the $TR$ technique and the $\mathrm{[S\,II]}(6717/6731)$ ratio are shown as solid lines and $1\sigma$-uncertainty shaded regions  (labelled). Beyond a radius of $r>4.6$\;arcseconds, there was insufficient signal in individual pixel rows for measurement of the [S\,II] lines, therefore we report the spatially-integrated value for all rows beyond this radius.}
	\label{fig: ne_sii_rows}
\end{figure}

To check that the density estimates measured for the narrow component in our main spectroscopic aperture are representative of non-outflowing gas in the NLR, we also measured the traditional [S\;II] and [O\;II] diagnostic ratios for individual pixel rows to the northeast of the nucleus --- the results are shown in Figure \ref{fig: ne_sii_rows}. Note that, to be considered a valid density estimate, we required the measured ratios to be at least 3$\sigma$ away from the high and low density limits as set by atomic physics ($0.41<\mathrm{[O\;II](3729/3726)}<1.50$; $3.01<\mathrm{[S\;II](4068/4076)}<3.28$, $0.46<\mathrm{[S\;II](6717/6731)}<1.45$; determined using the \textsc{PyNeb Python} package). Moreover, beyond a radial distance from the nucleus of $r>4.6$\;arcseconds, there was insufficient signal-to-noise to allow accurate measurement of  the $\mathrm{[S\,II]}\lambda\lambda6717,6731$ lines; therefore, we summed all flux values in pixel rows beyond this radius (and added the flux uncertainties in quadrature) to provide a single measurement. Clearly the narrow component -- assumed here to present the non-outflowing gas in the disk of NGC\,1068 -- maintains a low density of  $n_{e, \mathrm{[S\;II]}}\; \sim 250 - 350\,\mathrm{cm}^{-3}$
across the full radial range $2.8 < r <  6$\,arcseconds ($176 < r < 380$\,pc) over which it is detected.

Several recent studies have emphasised that the  traditional $\mathrm{[S\,II]}(6717/6731)$ and $\mathrm{[O\,II]}(3729/3726)$ ratios are not sensitive to higher electron densities \citep{Holt2011,Rose2018,Davies2020,HoldenTadhunter2023,Holden2026}. Therefore, we also measured densities  using the transauroral-line-ratio ($TR$) technique developed by \citet{Holt2011}, which involves comparing measured emission-line flux ratios to the predictions of photoionisation modelling; this method has the key advantage of being sensitive to a wide range of electron densities ($2.00<\mathrm{log}_{10}(n_e\;/\;\mathrm{cm}^{-3})<6.00$), and requires the measurement of the total flux of emission-line doublets, rather than individual fluxes of lines within closely-spaced doublets; it also produces estimates of the emission-line reddening. 

\begin{figure}
	\centering
	\includegraphics[width=\linewidth]{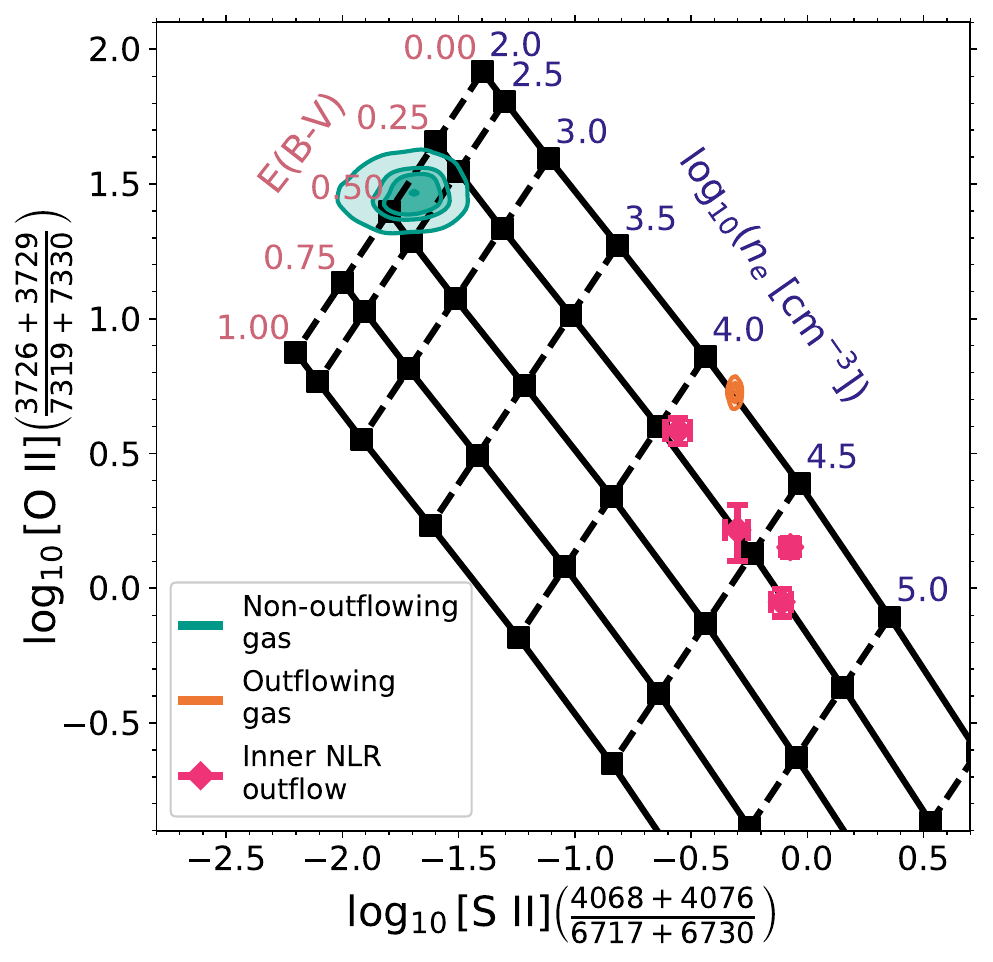}
	\caption{Modelled transauroral-line-ratio ($TR$) grid for solar-metallicity, dust-free gas with an ionisation parameter of $\mathrm{log}_{10}U=-3.00$, ionised by a central source of spectral index $\alpha=2.0$ (black grid); lines of constant reddening are solid, while lines of constant electron density are dashed. The probability density functions of the ratio values for the non-outflowing and outflowing components, as measured for our spectroscopic aperture, are shown in green and orange, respectively (with contours at the 16th, 50th, 64th and 99th percentiles). Moreover, we plot the values for the outflowing gas in the inner ($r <150$\;pc) NLR of NGC\;1068 (pink diamonds), which were presented by \citet{HoldenTadhunter2023}}
	\label{fig: tr_grid}
\end{figure}

In order to properly account for the uncertainties, we followed the Monte Carlo methodology presented in \citet{Holden2026} by randomly drawing 10,000 line fluxes from normal distributions, the mean and widths of which were taken to be the measured flux values and the $1\sigma$ uncertainties that we derived from our emission-line fits. For each of the 10,000 realisations, we calculated the line-flux ratios
\begin{align}
	& TR(\mathrm{[O\,II]}) = F(3726 + 3729)/F(7319 + 7331), \label{eq: tr_oii}\\
	& TR(\mathrm{[S\,II]}) = F(4068 + 4076)/F(6717 + 6731) \label{eq: tr_sii}
\end{align}
for the broad and narrow components. Then, the analytical expression derived by \citet{Holden2026} was used to calculate electron densities, taking the expression's constants as those corresponding to plane-parallel, radiation-bounded photoionisation for solar-metallicity, dust-free gas with a power-law ionising-continuum between 0.025\;nm -- 10\;\textmu m following $F_\nu\propto\nu^{-2.0}$ and an ionisation parameter of $\mathrm{log}_{10}U=-3.00$ (see Table C1 of \citealt{Holden2026}). This choice was based on the results of photoionisation modelling for the NLRs of nearby AGN, including NGC\;1068 \citep{Ferland1983, Baron2019b, Revalski2021, HoldenTadhunter2023}\footnote{Note that the results are not strongly sensitive to the choice
of photoionization model parameters \citep{HoldenTadhunter2023}}. The resulting $TR(\mathrm{[O\,II]})$ and $TR(\mathrm{[S\,II]})$ distributions from the 10,000 Monte Carlo realisations are presented along with a photoionisation grid of these parameters with varying electron density and extinction values (following the CCM89 $R_v=3.1$ extinction law) in Figure\;\ref{fig: tr_grid}. Using the equation from \citet{Holden2026} to calculate a density for each realisation, and then taking the 50th percentiles of the overall distribution as the parameter value (and estimating $1\sigma$ uncertainties from the 16th and 84th percentiles), we found that the electron densities for the narrow and broad components are 
$n_{e, \mathrm{TR}}\;=128^{+63}_{-30}\,\mathrm{cm}^{-3}$  and $n_{e, \mathrm{TR}}\;=14\,125^{+666}_{-635}\,\mathrm{cm}^{-3}$, respectively, corresponding to a density contrast of $110^{+49}_{-26}$. Therefore, regardless of whether we use the traditional density diagnostics or the transauroral method, we measure a high density contrast between the broad (outflowing) and narrow (non-outflowing) components.

In comparison with estimates of the NLR densities from previous studies of NGC\,1068 that did not separate outflow and non-outflow components, the high electron densities that we measure for the outflowing gas with the $TR$ technique are consistent with those found using both the same method \citep{HoldenTadhunter2023} and detailed photoionisation modelling \citep{Revalski2021, Revalski2022} on smaller radial scales ($r<150$\;pc). Also, a recent JWST study derived similarly high electron densities for the high-ionisation coronal-line gas in NLR using mid-infrared emission-line ratios of [Ne\;V] and [Ar\;V] \citep{Marconcini2025}. 

In addition to the electron densities, by comparing the measured ratios to a finer $TR$ grid of simulated points than is shown in Figure\;\ref{fig: tr_grid}, we found that the non-outflowing gas has $TR$ values consistent with zero reddening, while the outflowing gas has a measured value of $E(B-V)_{TR}=0.42^{+0.20}_{-0.19}$. These results are consistent with those derived by comparing the H$\gamma$/H$\beta$ and H$\delta$/H$\beta$ Balmer-line ratios with the theoretical Case B recombination values of 0.466 and 0.256, respectively (assuming a temperature of $T=10,000$\;K: \citealt{Osterbrock2006}): $E(B-V)_{\mathrm{H}\delta/\mathrm{H}\beta} = 0.20\pm0.02$ and $E(B-V)_\mathrm{H\gamma/H\beta}=0.22\pm0.02$ for outflowing component; and $E(B-V)_{\mathrm{H}\delta/\mathrm{H}\beta} = 0.00\pm0.01$ and $E(B-V)_{\mathrm{H}\gamma/\mathrm{H}\beta} = 0.00\pm0.01$ for the non-outflow component. The Balmer-derived values for the non-outflowing gas are within $1\sigma$ of that derived from the transauroral-line grid, while those for the outflowing gas imply that this component suffers negligible dust extinction.

\subsection{Dust-depletion diagnostics}
\label{section: results: dust_destruction}

\subsubsection{Coronal-line measurements for the high-ionisation gas}
\label{section: results: dust_destruction: high_ionisation}

In line with previous optical and infrared spectroscopic observations of the inner NLR of NGC\;1068 ($r < 150$\:pc: \citealt{Crenshaw2000_N1068,May2017,HoldenTadhunter2023}), prominent [Ne\,V] and [Fe\,VII] coronal lines are detected in our extracted spectrum (Figure\;\ref{fig: radio_lobe_spectrum_full}). Since neon is a noble gas and therefore not depleted onto dust grains in the ISM, whereas iron is a refractory element and heavily depleted \citep{Jones2007}, the $\mathrm{[Ne\,V]}\lambda3426/\mathrm{[Fe\,VII]}\lambda6087$ ratio is strongly sensitive to different levels of depletion of refractory elements onto dust grains. Moreover, since Ne$^{4+}$ and Fe$^{6+}$ have similar ionisation potentials ($E_\mathrm{ion} \sim 100$\,eV) and the relevant emission lines have similar critical densities, their ratio is relatively insensitive to the ionisation conditions and density. In this context, it is striking that the non-outflow component has a $\mathrm{[Ne\,V]}\lambda3426/\mathrm{[Fe\,VII]}\lambda6087$ ratio that is a factor of 14.4$\pm$2.9 times higher than that of the outflow, whose small $\mathrm{[Ne\,V]}\lambda3426/\mathrm{[Fe\,VII]}\lambda6087$ ratio is consistent with those measured for the warm outflows on smaller radial scales using HST data ($r < 150$\,pc; \citealt{HoldenTadhunter2023}). The difference in the relative strengths of the outflow and non-outflow components between the $\mathrm{[Ne\,V]}\lambda3426$ and $\mathrm{[Fe\,VII]}\lambda6087$ lines is clearly apparent in the line profiles shown in Figure\;\ref{fig: spectral_fits}.

\begin{figure}
	\centering
	\includegraphics[width=1\linewidth]{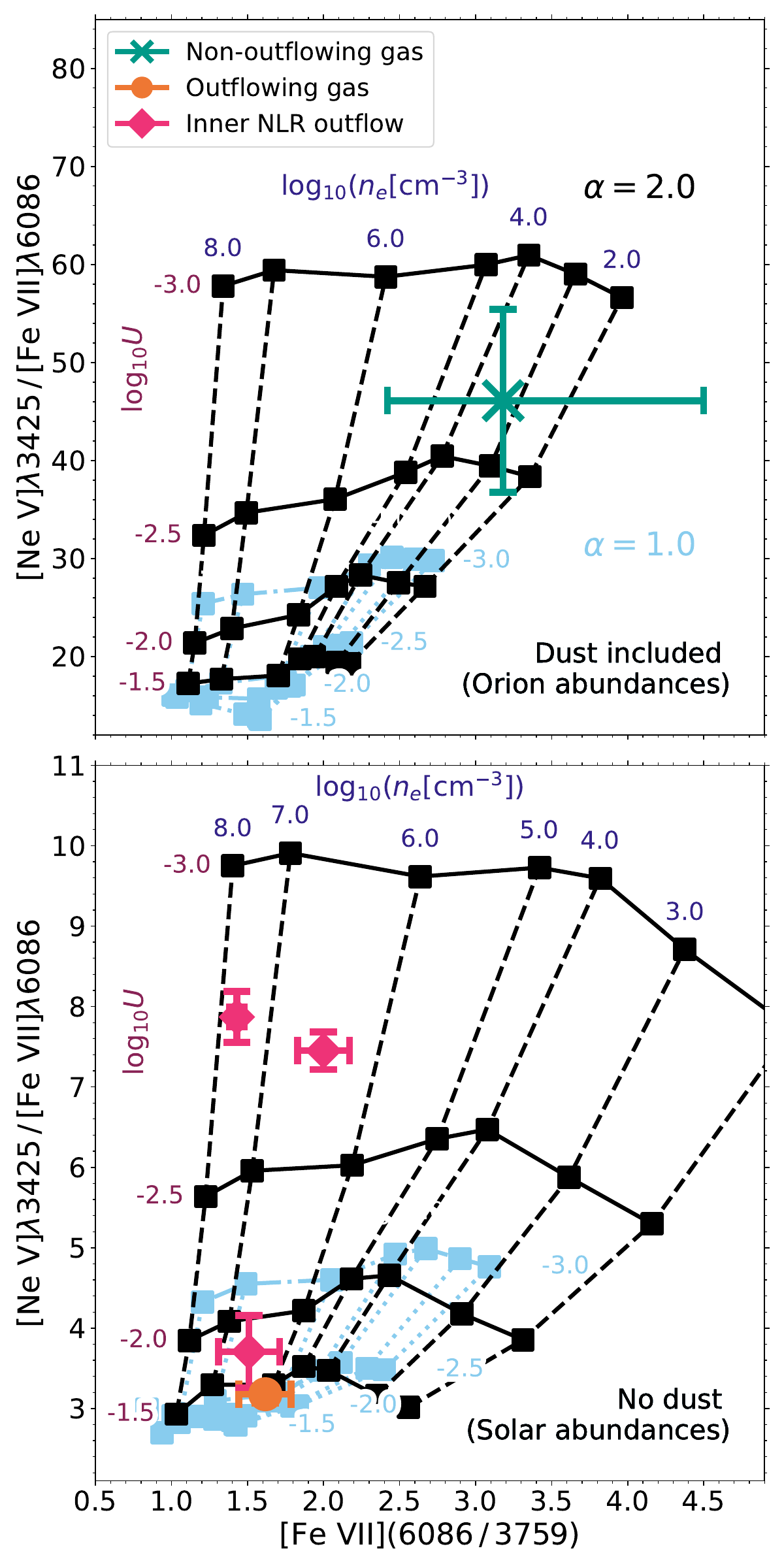}
	\caption{$\mathrm{[Fe\,VII]}(6086/3759)$ vs $\mathrm{[Ne\,V]}\lambda3425/\mathrm{[Fe\,VII]}\lambda6086$ diagnostic diagram for the high-ionisation coronal gas ($E_\mathrm{ion}>100$\;eV) at the base of the north-eastern radio lobe in NGC\;1068. The grids in the top panel show the predicted values of AGN photoionisation models for gas that contains dust grains, while the grids in the bottom panel are for dustless, solar-abundance gas. The black grids are for an AGN power-law continuum spectral index of $\alpha=2$ (assuming $F_{\nu} \propto \nu^{-\alpha}$, as
    labelled), with solid lines of constant ionisation parameter and dashed lines of constant electron density (both labelled); the blue grids are for $\alpha=2$, and the constant ionisation parameter and electron density lines are dotted and dash-dotted, respectively. The measured values for the extracted spectrum are shown (orange circle: outflowing gas; green cross: non-outflowing gas). In addition, we plot the line ratios presented by \citet{HoldenTadhunter2023} for the outflowing-gas emission in the inner ($r<150$\;pc) part of NGC\;1068's NLR (pink diamonds).}
	\label{fig: fevii_fevii_nev_grid}
\end{figure}

To investigate the difference in the coronal lines ratios between the outflow and non-outflow components, we first used version 23 of \textsc{Cloudy} \citep{Chatzikos2023} to generate radiation-bounded photoionisation models of varying electron density and ionisation parameter with 0.025\;nm -- 10\;\textmu m spectral indices of $\alpha=2.0$ and $\alpha=1.0$ (where $F_\nu\propto\nu^{-\alpha}$), chosen to cover the range of spectral indices typically deduced from photoionisation modelling of samples of AGN. The modelling results are shown in Figure\;\ref{fig: fevii_fevii_nev_grid}); we note that assuming a different spectral index or assuming a different form of ionising-continuum (including a black-body of temperature $T=10^5$\;K) does not  change our main results or conclusions. We produced these models for two abundance sets: solar-abundance gas with no depletion of elements onto dust grains (shown as the grid in the bottom panel of Figure\;\ref{fig: fevii_fevii_nev_grid}) and H\,II-region abundances with dust-depletion (top panel of Figure\;\ref{fig: fevii_fevii_nev_grid}). The latter are a subjective mean of the abundances for the Orion nebula \citep{Baldwin1991, Rubin1991, Rubin1993, Osterbrock1992}, with large-R type grains \citep{Baldwin1991}; this abundance set is provided with the \textsc{Cloudy} code as `HII.abn'. We then used the simulated values of the $\mathrm{[Fe\,VII]}(6086/3759)$ and $\mathrm{[Ne\,V]}\lambda3425/\mathrm{[Fe\,VII]}\lambda3759$ flux ratios to produce a diagnostic diagram that is sensitive to the dust content, electron density, and ionisation parameter ($U$) of the gas \citep{Rose2011}; the presence of dust in the models not only depletes iron atoms, but also changes the ionisation balance of the gas clouds \citep{McKaig2024}. The extinction value of the simulated dusty gas cloud, $A_{\rm v}=0.66$, is consistent with our estimates for the non-outflowing gas obtained using Balmer decrement ($A_{\rm v} =0.65\pm0.09$) and transauroral-line ratio ($A_{\rm v}=1.30^{+0.62}_{-0.59}$) methods, where we have assumed a total-to-selective extinction ratio of $A_{\rm v} / E(B-V) = 3.1$ when converting from our $E(B-V)$ measurements.

Comparing these grids to the ratio values measured from our spectrum reveals that the $\mathrm{[Ne\,V]}\lambda3425/\mathrm{[Fe\,VII]}\lambda6086$ ratio for the outflow component is consistent with the dust-free models, while the much higher value measured for the non-outflow component clearly requires ISM levels of dust depletion. Hence, given the relatively small variations in the photoionisation model due to different assumed values of the ionisation parameter, shape of the ionising continuum, and electron density, the large difference in this emission-line flux ratio between the components can only be explained in terms of contrasting neon to iron abundance ratios, most likely due to different levels of depletion of iron onto dust.

The reddening values that we obtained as a by-product of the transauroral-line technique for measuring the electron density, and also from the H$\delta$/H$\beta$ and H$\gamma$/H$\beta$ Balmer-line ratios, further reinforce this result: as we showed in Section \ref{section: results: electron_density_measurements}, the outflowing gas has values consistent with zero reddening due to dust, while the non-outflowing gas presents colour-excess values of $E(B-V)_{TR}=0.42^{+0.20}_{-0.19}$ and $E(B-V)_\mathrm{Balmer}=0.21\pm0.03$, respectively, for the two methods. This is further evidence that the non-outflowing gas contains a significant amount of dust, while the outflow is largely dust-free.

\subsubsection{Dust depletion in the low-ionisation gas}
\label{section: results: dust_destruction: low_ionisation}

\begin{figure*}
	\centering
	\includegraphics[width=\linewidth]{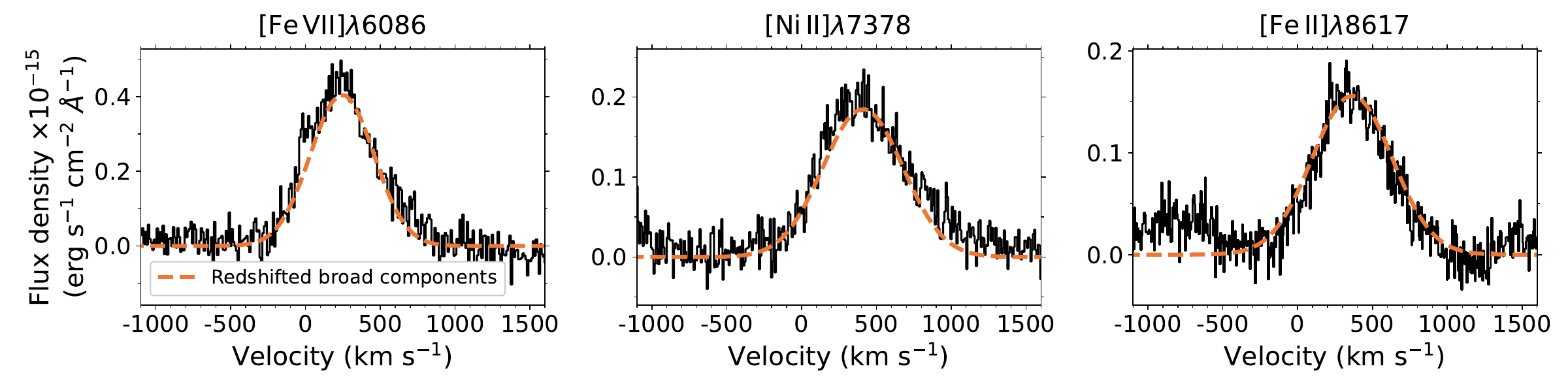}
    \vspace{-0.4cm}

    \caption{Comparison of the velocity profiles of a high-ionisation emission-line ($E_\mathrm{ion}\sim100$\;eV; left panel) and low-ionisation emission lines ($E_\mathrm{ion}\sim8$\;eV; middle and right panels) of the refractory elements iron and nickel, as seen in the spectrum extracted from our Xshooter data (see Figure\;\ref{fig: slit_position}). The redshifted broad components from our spectral fits (Figure\;\ref{fig: spectral_fits}) --- which we interpret as arising from outflowing gas --- are shown as dashed orange lines for reference.}
	\label{fig: refractory_velocity_profiles}
\end{figure*}

\begin{figure}
	\centering
	\includegraphics[width=1\linewidth]{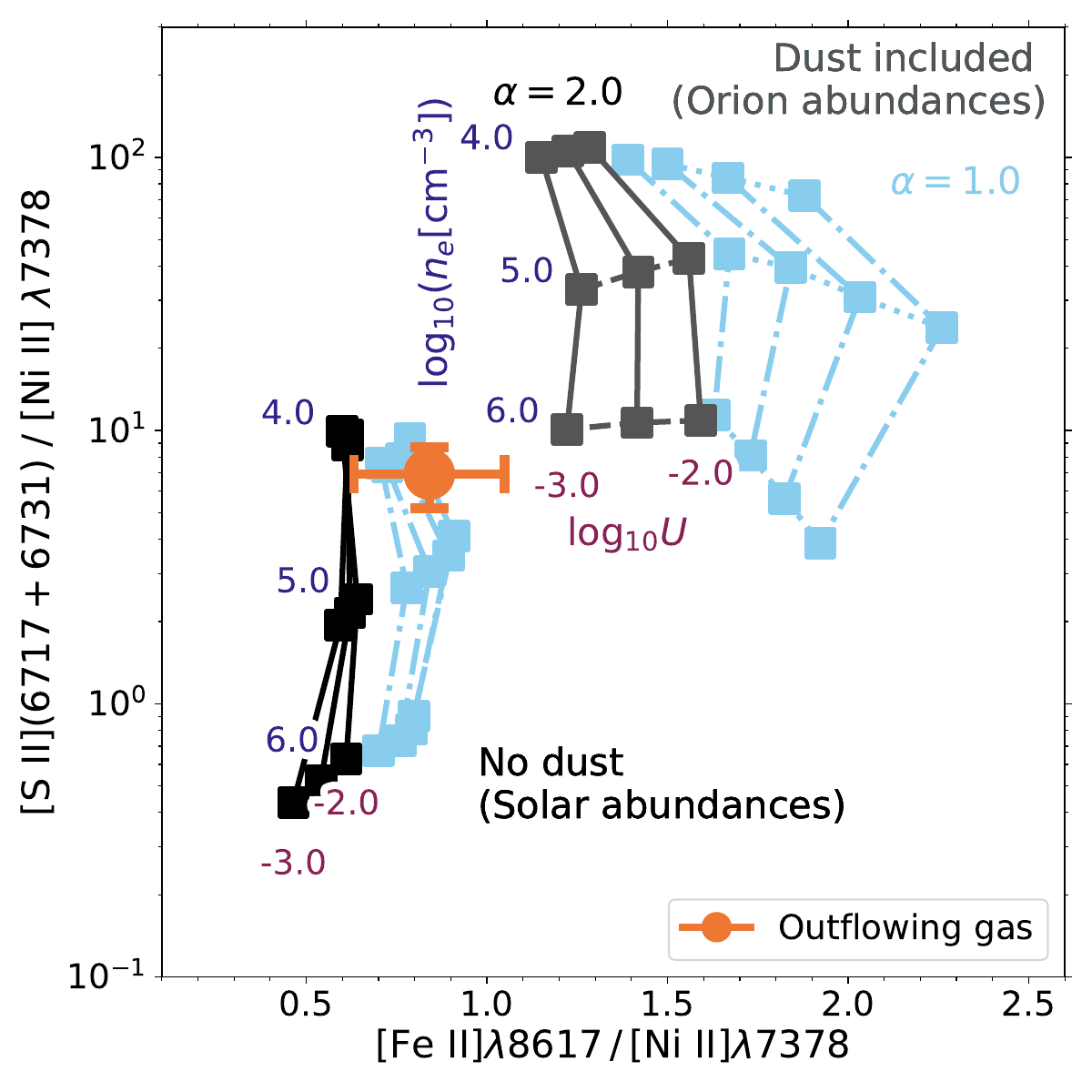}
	\caption{$[\mathrm{Fe \ II}] \lambda 8617/[\mathrm{Ni \ II}]\lambda 7378$ vs $[\mathrm{S\ II}](6717 + 6731)/\mathrm{[Ni \ II}]\ \lambda 7378$ diagnostic diagram for low-ionisation gas ($E_\mathrm{ion}\sim8$\;eV), with the ratios measured for
    our spectroscopic aperture indicated by the orange symbol. The grid and marker schemes are the same as that of the high-ionisation counterpart to this diagnostic diagram (Figure\;\ref{fig: fevii_fevii_nev_grid}), however note here that the y-axis is logarithmic.}
	\label{fig: ni2_s2_fe2_grid}
\end{figure}

It is notable that, as well as the high-ionisation [Fe\;VII] lines,  lower-ionisation forbidden emission lines of refractory elements are also detected in the spectrum, including [Fe\;II]$\lambda$8617, [Fe\;III]$\lambda$4658, [Fe\;V]$\lambda$5167, [Fe\;VI]$\lambda$5146 and [Ni\;II]$\lambda$7378. Significantly, none of these lines show evidence for narrow components emitted by the non-outflowing gas; all are dominated by a broad, redshifted component (see Figure \ref{fig: refractory_velocity_profiles}). To investigate the impact of different levels of dust depletion on the relative strengths of these lower ionisation lines of refractory elements, we again used \textsc{cloudy} \citep{Ferland2017} to create a $[\mathrm{Fe \ II}] \lambda 8617/[\mathrm{Ni \ II}]\lambda 7378$ vs $[\mathrm{S\ II}](6717 + 6731)/\mathrm{[Ni \ II}]\ \lambda 7378$ diagnostic diagram, which is shown in Figure\;\ref{fig: ni2_s2_fe2_grid} --- this may be considered  a low-ionisation counterpart to the $\mathrm{[Fe\,VII]}(6086/3759)$ vs $\mathrm{[Ne\,V]}\lambda3425/\mathrm{[Fe\,VII]}\lambda6086$ diagram presented in Figure\;\ref{fig: fevii_fevii_nev_grid}. 

Although the separation between the dusty and dust-free models is not as distinct as for the diagnostic diagram involving the high-ionisation coronal line ratios --- an issue already discussed extensively for the [Fe\;II] lines detected in AGN spectra at near-infrared wavelengths \citep[e.g.][]{Simpson1996} --- the outflow component has line ratios that are more consistent with those predicted for dust-free gas, assuming that the electron density is the same as that derived from the transauroral [O\;II] and [S\;II] lines ($n_e \sim 10^{4.15}$\,cm$^{-3}$); photoionisation models that include ISM-levels of dust are only consistent with the measured [Ni\;II]/[S\;II] ratio at much higher electron densities. Considering both these results and those for the coronal-line ratios, this implies that all the outflowing gas --- whether of low or high ionisation --- shows evidence for significant levels of depletion of refractory elements onto dust grains.

\subsubsection{The non-detection of $\mathrm{[Ca\,II]}\lambda7291$}
\label{section: results: dust_destruction: caII}

\begin{figure}
	\centering
	\includegraphics[width=\linewidth]{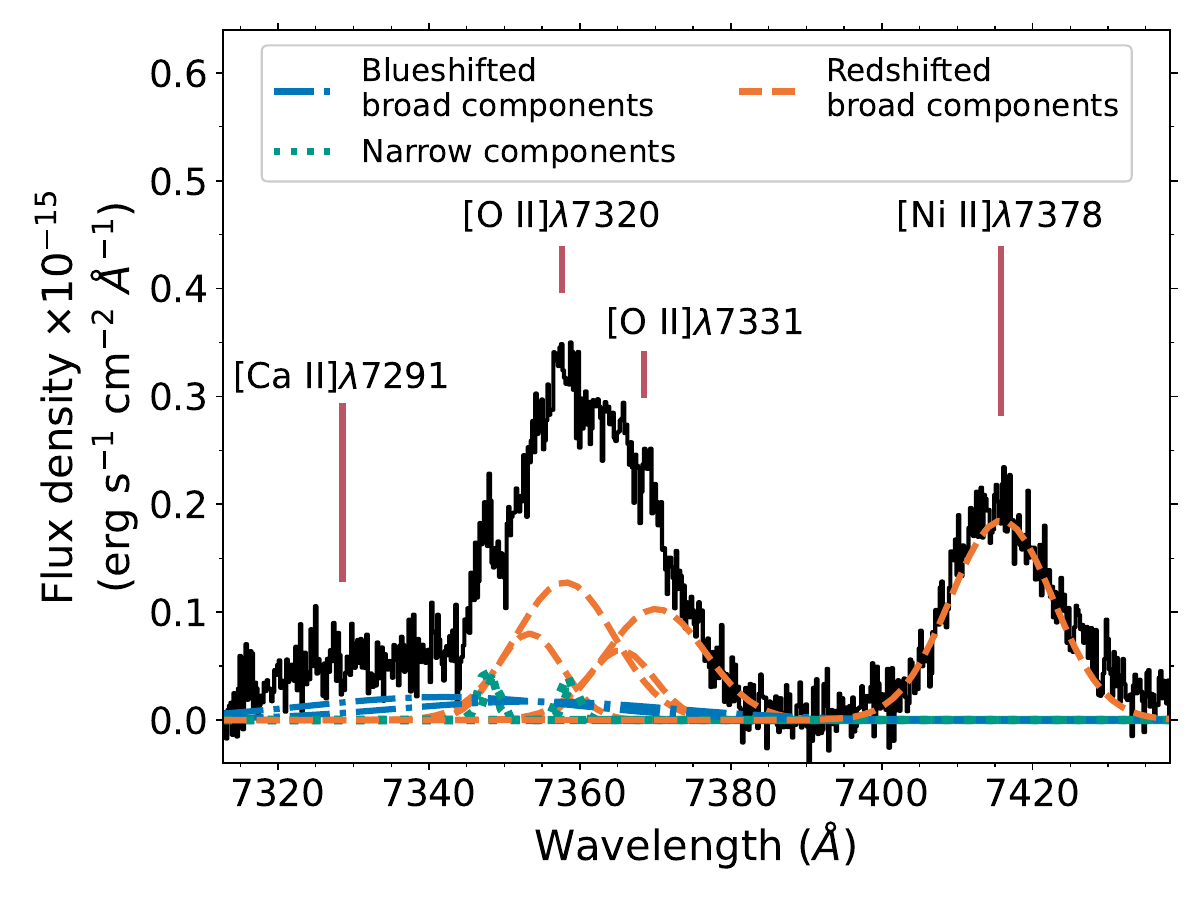}
    \vspace{-0.4cm}
	\caption{A spectrum extracted from a region at the base of the northeastern radio lobe in NGC\;1068 in the wavelength region of the $\mathrm{[Ca\,II]}\lambda7291$ line, which is not detected. Other lines, including the refractory line $\mathrm{[Ni\,II]}\lambda7378$ that has a similar ionisation energy and critical density, are shown for comparison. The labels shown by the red vertical lines are positioned at a velocity shift of 400\;km\;s$^{-1}$, corresponding to the approximate centre of the redshifted outflow component (Figure\;\ref{fig: refractory_velocity_profiles}).}
	\label{fig: caii_7291_spectral_region}
\end{figure}

A key low-ionisation refractory-element line that we do not detect in our spectrum is $\mathrm{[Ca\,II]}\lambda7291$ (see Figure \ref{fig: caii_7291_spectral_region}), which has a similar ionisation potential and critical density to the other low-ionisation lines we considered in Section \ref{section: results: dust_destruction: low_ionisation}. The absence of this line --- which should be relatively strong based on the predictions of AGN photoionisation models for dust-free gas \citep{Ferland93} --- has been used to argue that the NLR clouds are dusty \citep[e.g.][]{Villar96,Villar97,Shields1999b}. Since we were unable to measure the $\mathrm{[Ca\,II]}\lambda7291$ line flux directly, we forced a spectral-fitting process at the wavelength of the putative redshifted  [Ca\,II] line by assuming that it has a profile identical in shape  to that of $\mathrm{[Ni\;II]}\lambda7378$, in order to provide a $3\sigma$ upper limit on the line flux. We then produced a diagnostic diagram consisting of $\mathrm{[Ca\,II]}\lambda7291$/$\mathrm{[O\,II]}(7319\lambda + 7330\lambda)$ vs $\mathrm{[Ca\,II]}\lambda7291$ / $\mathrm{[Ni\,II]}\lambda7378$, and plotted the measured upper limits.

\begin{figure}
	\centering
	\includegraphics[width=\linewidth]{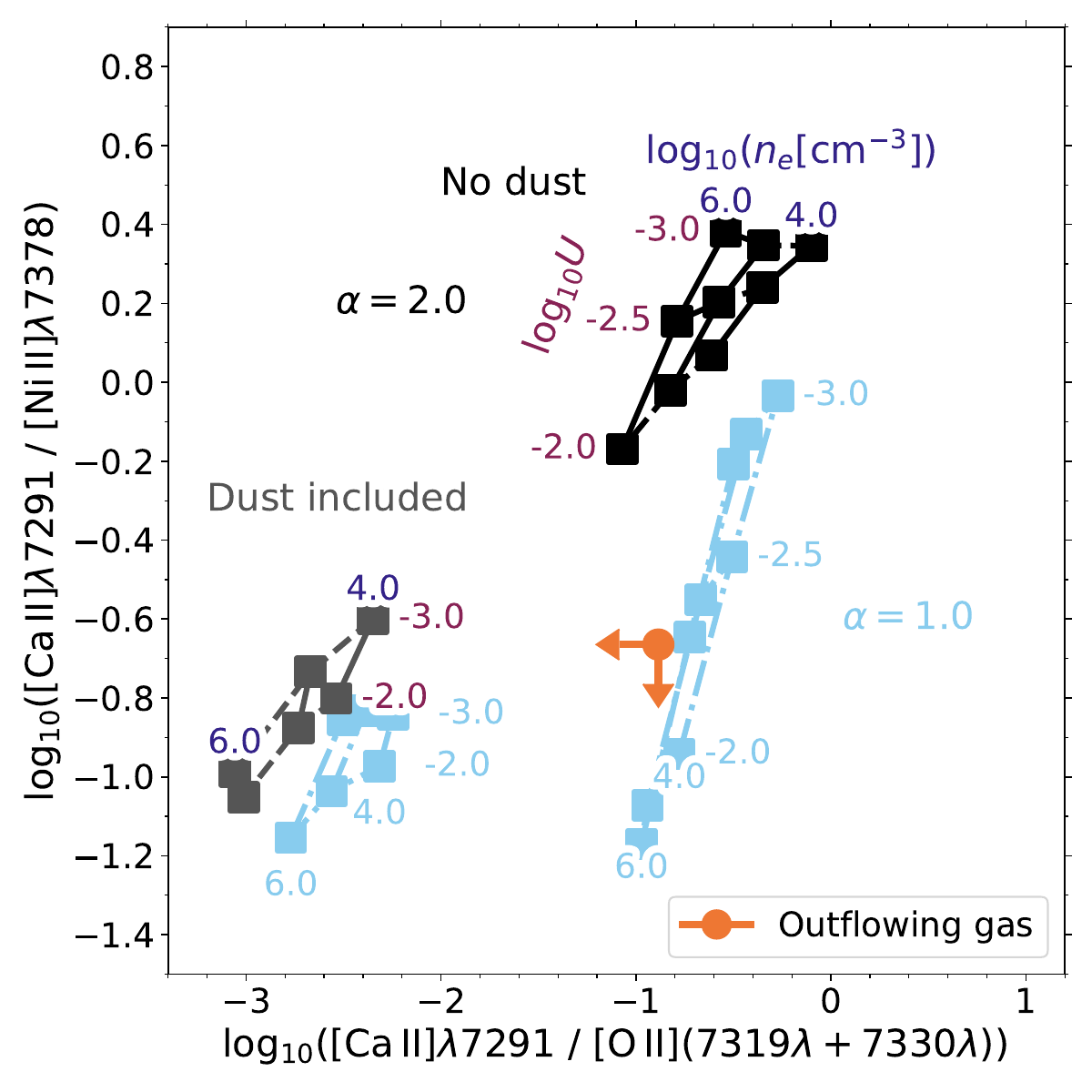}
    \vspace{-0.4cm}
    \caption{Diagnostic diagram consisting of the $\mathrm{[Ni\,II]}\lambda7378$ / $\mathrm{[O\,II]}(7319\lambda + 7330\lambda)$ vs $\mathrm{[Ca\,II]}\lambda7291$ / $\mathrm{[Ni\,II]}\lambda7378$ emission-line ratios, with photoionisation grids shown with the same convention as in Figures\;\ref{fig: fevii_fevii_nev_grid} and \ref{fig: ni2_s2_fe2_grid}. Note that, due to $\mathrm{[Ca\,II]}\lambda7291$ not being detected in our spectrum, the measured ratio values are upper limits for the broad component.}
	\label{fig: caii_ni_oii_grid}
\end{figure}

The results are presented in Figure\;\ref{fig: caii_ni_oii_grid}, from which it can be seen that dustless gas can explain our measured upper ratio limits only for models that combine a relatively hard ionising continuum
($\alpha=1$) with a high ionisation parameter (log$U=-2.0$) and a high electron density ($n_e=10^{6.0}$\;cm$^{-3}$); otherwise, gas with significant dust content is required. Since the latter case can more readily explain our observed line-ratio upper limits, we favour the interpretation the outflow must have significant dust content in order to explain the non-detection of $\mathrm{[Ca\,II]}\lambda7291$. This would imply that, unlike other refractory elements, much of the calcium in the outflowing gas remains locked in dust grains.

\section{Discussion}
\label{section: discussion}

To investigate the NLR outflows in NGC\,1068, we have  used a range of emission-line diagnostic ratios measured from VLT/Xshooter spectra of a region at the base of its northeastern radio lobe. In this section, we discuss the results in the context of outflow acceleration mechanisms, the degree of dust destruction, the utility of coronal lines as shock diagnostics, and the recycling of the ISM in the narrow-line region.

\subsection{Evidence for shock-accelerated outflows in NGC\;1068}
\label{section: discussion: shocks}

In order to determine the dominant outflow acceleration mechanism, it is important to compare our results regarding gas compression and refractory-element depletion with the predictions of the radiation-pressure and fast-shock models.

To investigate whether the high gas densities we measure for the outflow could be due to the clouds being radiation-pressure confined, we used the models of \citet{McKaig2024} to predict the electron densities of a radiation-pressure-confined (RPC) cloud at the location of the ionisation front, where the key [S\;II] and [O\;II] diagnostic emission lines are expected to be produced. Using Equation\;5 of that study for dust-free gas with an estimated bolometric luminosity for NGC\;1068 of $L_\mathrm{bol}=10^{45}$\;erg\;s$^{-1}$ \citep{Woo2002, LopezRodriguez2018, Gravity2020}, assuming a typical ionising-to-total luminosity ratio of 0.3 \citep{Elvis1994, Lusso2015}, and taking the cloud distance from the AGN to be the observed (projected) distance of the spectroscopic aperture (230\;pc), we derive a ionisation-front gas density of 500\;cm$^{-3}$; this density would be significantly lower (by a factor of up to 1.7) if we corrected the distance for projection. 

Critically, the RPC-predicted density value is factors of 11 and 28 lower than the densities we measure for the outflow component using the $\mathrm{[S\;II]}(6717/6731)$ ratio and transauroral methods respectively (Section\;\ref{section: results: electron_density_measurements}), indicating that radiation pressure alone cannot explain the high densities we measure for the outflow. On the other hand, compression factors of $\sim$100 or more, and correspondingly high densities for the outflow component, are possible for gas cooling in pressure equilibrium behind a shock front \citep{Sutherland2017}. Therefore, our measured densities provide clear evidence for shocks compressing the gas at the base of the NE radio lobe in NGC\;1068.

In terms of dust properties, the contrast between the outflow and non-outflow components is likely to be due to the outflow acceleration mechanism destroying the dust. At the 230\;pc radial distance of our spectroscopic aperture from the nucleus, the incident radiation from the AGN alone is insufficient to destroy dust by heating it to its sublimation temperature; such radiative dust destruction is important only on sub-parsec scales in the inner part of the circum-nuclear torus that obscures the AGN from our direct view \citep{Gravity2020}\footnote{Note that the most recent radiative acceleration model for the outflows in NGC\,1068 \citep{Meena2023} requires that the outflowing gas currently observed at the location of our spectroscopic aperture originated at radii between 10 and 25\,pc from the nucleus -- well outside the dust sublimation radius.}. Instead, the dust in the outflow is likely to have been destroyed by sputtering --- the erosion of dust grains due to collisions with fast, ionised gas particles  --- in fast shocks \citep[e.g.][]{DeYoung1998,VillarMartin2001} that also compress the outflowing gas and accelerate it to the high velocities that we observe. Following this shock acceleration, the largely dust-free gas in the outflow has cooled and, like the non-outflow component, is currently photoionised by the AGN (Section\;\ref{section: results}; Table\;\ref{tab: line_ratios}). Together with the evidence for gas compression, the contrast between the dust properties of the outflow and non-outflow components provides strong evidence that the warm outflows in NGC\,1068 have been accelerated by fast shocks. 

\subsection{Is all the dust destroyed in the shocks?}
\label{section: discussion: caii}

Considering the evidence that shocks destroy dust grains during the outflow acceleration process, it is interesting that we do not detect the $\mathrm{[Ca\,II]}\lambda7291$ line in our spectrum, given that dust-free AGN photoionisation models suggest that it should be relatively strong compared to other low-ionisation lines (see Section\;\ref{section: results: dust_destruction: caII}; Figures\;\ref{fig: caii_7291_spectral_region} and \;\ref{fig: caii_ni_oii_grid}). Clues to the reason for this may be provided by the fact that calcium is a super-refractory element, in the sense that it has a higher dust-grain condensation temperature ($T_c\sim1500$\;K) than iron and nickel ($T_c\sim1300$\;K: \citealt{Wood2019}), leading to it forming the cores of dust grains and thus being significantly more heavily depleted in the ISM \citep{Jenkins1989, Crinklaw1994}. 

As dust grains pass through shocks, their mantles (which contain iron and nickel) must first be destroyed before the calcium in the super-refractory grain cores can be liberated \citep{Clayton1982, Jones1994, Bocchio2014}. Therefore, a possible explanation for the absence of the $\mathrm{[Ca\,II]}\lambda7291$ line in our NGC 1068 spectrum, despite iron and nickel lines of similar ionisation and critical density being detected, is that, while the dust grain mantles have been destroyed in shocks, the super-refractory cores have survived. In this scenario, much of the calcium may remain locked in grains post-shock, preventing significant $\mathrm{[Ca\,II]}\lambda7291$ emission. Alternatively, the calcium may already have been deposited in the cores of dust grains that are re-forming in the post-shock gas, but the iron-rich mantles of the dust grains have yet to re-form. 

Although ALMA 349\;GHz imaging by \citet{GarciaBurillo2014} indicates that there is little (or no) cold-dust emission between the AGN and an arc structure 4--7\;arcsec (250--440\;pc) from the nucleus that coincides with the NE radio lobe ($r\sim390$\;pc), we note that there is evidence for the presence of \textit{some} warmer dust on these scales. This is provided by the detection of extended mid-IR continuum emission at 12.8\,$\mu$m that is closely aligned with the extended radio structure of NGC\;1068 \citep{Alloin2000,Galliano2005}; recent JWST results have suggested that the warm dust in such aligned structures may be heated by jet-induced shocks \citep{Haidar2026}.

This analysis demonstrates the value of comparing diagnostic ratios involving emission lines of refractory elements with different levels of depletion in the ISM. With such an approach, we can not only learn about the degree of dust destruction, but potentially also derive information on the structure of the dust grains (such as the existence of super-refractory cores).

\subsection{Coronal lines as tracers of shocks and AGN feedback}
\label{section: discussion: cls}

The results of this study inform a growing literature that points to the coronal lines as important shock diagnostics. Notably, based on their photoionisation modelling work, \citet{McKaig2024} have emphasised that having gas with low levels of depletion of refractory elements onto dust grains is key for the observation of strong coronal lines of iron \citep[see also][]{Mullaney2009,Lamperti2026}; other factors that affect their strength include the shape of the ionising continuum, the ionisation parameter, and the electron density \citep[e.g.][]{Taylor2003}. Indirect evidence for coronal line production as a result of shocks is provided by several examples of alignments of spatially-extended coronal line emission with the radio-emitting jets of relativistic particles associated with some AGN \citep{Tadhunter1988,MullerSanchez2011,FonsecaFaria2023,RodriguezArdila2025b}. Here, by comparing the outflow and non-outflow gas properties in NGC\,1068, we have directly confirmed the importance of fast shocks as an acceleration mechanism in the NLR, and of iron coronal lines as a key diagnostic of such shocks (Figure\;\ref{fig: fevii_fevii_nev_grid}).

Provided that the ionisation conditions are suitable (i.e. high ionisation parameter, high gas density and/or a hard ionising continuum) we may distinguish two main circumstances under which strong coronal line emission of refractory elements is likely to be produced, both involving dust destruction: the unresolved innermost part of the NLR ($r < 10$\;pc, perhaps the inner face of the circumnuclear torus: \citealt{Murayama1998,Rose2011}), where the flux of radiation from the AGN is sufficient to destroy the dust by heating it to the sublimation temperature; and the more extended NLR ($r\sim0.01$--5\;kpc), where jet- or wind-driven shocks may destroy the dust by sputtering and compress the gas, as is seen in NGC\,1068. 

\subsection{The recycling of the ISM in shocks}
\label{section: discussion: cls}

Our findings are also important for understanding the re-cycling of the interstellar medium in the NLR, and the multi-phase nature of the AGN-induced outflows. There is now considerable evidence that AGN outflows exist across a range of gas phases from hot ($10^6$ -- $10^8$\,K) to warm ($\sim$10$^4$\,K) to cold ($<$100\,K), and that the cold molecular phase may dominate the mass and kinetic power budgets of the outflows \citep{Fiore2017, Speranza2024, Holden2024}.
Interestingly, in the case of NGC\,1068, cold molecular outflows are detected across the inner parts of the NLR out to a radius of $\sim$\;400\;pc \citep{GarciaBurillo2014}, albeit with much lower velocities ($\Delta V \sim 50$ -- 100\,km s$^{-1}$) than measured for the warm ionised outflows. 

Since dust is essential for the efficient formation of molecules, the detection of molecular outflows may at first sight seem inconsistent with the evidence for dust destruction provided by the coronal lines. However, the contrast in their gas kinematics suggests that the molecular and warm outflows are spatially distinct, so that their acceleration mechanisms may not be the same. For example, rather than fast shocks, it is possible that some molecular material may be gently entrained in part of an accretion-disk wind \citep{GarciaBurillo2014}, and thus avoid dust and molecule destruction. However, it is notable that, based on observed intensity ratios between neutral-atomic and cold-molecular carbon monoxide emission in the NLR of NGC\;1068, \citet{Saito2022} argued that entrained molecular gas could be dissociated by shocks, thus forming the neutral atomic phase. Alternatively, the detection of mid-IR continuum emission in the extended NLR of this object \citep{Alloin2000,Galliano2005} along with non-detection of the [CaII]$\lambda$7291 line (Section\;\ref{section: results: dust_destruction: caII}) suggests that a small amount of dust may survive destruction in the shocks that accelerate the outflows, or re-form post-shock; this may allow the (re-)formation of molecules within the outflow as the gas cools.

\section{Conclusions}
\label{section: conclusions}

 We have analysed deep  VLT/Xshooter spectra of a region at the base of the northeastern radio lobe in NGC\;1068 to investigate the acceleration mechanism for its warm NLR outflows. In particular, we have contrasted the properties of the outflow with those of the  non-outflowing gas in the disk of the galaxy. Our conclusions are as follows.

\begin{itemize}
	\item A range of electron-density diagnostics --- including the traditional $\mathrm{[S\,II]}(6717/6731)$ ratio and the transauroral-line ratios (which are sensitive to a wider range of values) --- demonstrate that the outflowing gas in the NLR has a density 19--110 times larger than that of the non-outflowing gas in the galaxy's disk at the same projected radial distance from its nucleus.
	\item By developing new diagnostic diagrams involving the flux ratios of both high- and low-ionisation emission lines of refractory and non-refractory elements, we have shown that much --- if not all --- of the dust in outflowing gas has been destroyed, in contrast to the ISM amounts of dust present in the non-outflowing gas. This evidence for dust destruction is supported by the contrasting levels of dust extinction measured for the two kinematic components.
\end{itemize}
\vglue 0.1cm\
Taken together, these results imply that the warm NLR outflows in  NGC\,1068 are accelerated by fast shocks induced in the ISM by the AGN (either via a wind or a jet)  that both compress the gas and destroy the dust.

Clearly, the ratio of the coronal  $\mathrm{[Ne\,V]}\lambda3425$ and $\mathrm{[Fe\,VII]}\lambda6086$ lines is a particularly useful diagnostic of the importance of shocks in the NLR, since it involves lines of similar ionisation and critical density that are produced by refractory and non-refractory elements, the former of which are liberated from dust by shocks. However, the ratios of near- and mid-infrared lines such as $\mathrm{[Ne\,V]}\lambda\lambda14.3,24.3$\;$\mu$m, $\mathrm{[Si\,VI]}\lambda1.965$\;$\mu$m (silicon is a refractory element), and $\mathrm{[Fe\,VII]}\lambda\lambda7.81,9.52$\;$\mu$m \citep[e.g.][]{RamosAlmeida2025,Marconcini2025} may be used for a similar purpose, exploiting the high sensitivity and resolution of the James Webb Space Telescope. Future observations of larger samples of AGN will determine whether such coronal-line ratios are correlated with the emission-line kinematics, as expected if shock acceleration of the outflowing gas clouds is important in the spatially-resolved NLRs of the AGN population as a whole.

\section*{Acknowledgements}

The authors thank the anonymous referee for their helpful feedback. LRH and DJBS acknowledge support from the UK Science and Technology Facilities Council (STFC) in the form of grant ST/Y001028/1; DJBS acknowledges financial support from the UK's Leverhulme Trust via Research Project Grant RPG-2025-078; MIA acknowledges support from the STFC under grant ST/Y000951/1. MAB is supported by a UKRI Stephen Hawking Fellowship (EP/X04257X/1). IMM acknowledges support from Development in Africa with Radio Astronomy (DARA, Phase 3) through the UK’s Science and Technologies Facilities Council (STFC) grant ST/Y006100/1. Based on observations collected at the European Southern Observatory under ESO programme 111.24FW.001 (PI Holden). e-MERLIN is a National Facility operated by the University of Manchester at Jodrell Bank Observatory on behalf of STFC. The Karl G. Jansky Very Large Array (VLA) is operated by the National Radio Astronomy Observatory (NRAO); the NRAO and Green Bank Observatory are facilities of the U.S. National Science Foundation operated under cooperative agreement by Associated Universities, Inc. This research has made use of the NASA/IPAC Infrared Science Archive, which is funded by the National Aeronautics and Space Administration and operated by the California Institute of Technology. Photoionisation calculations were performed with version 23 of \textsc{Cloudy} (last described by \citealt{Chatzikos2023}). The authors thank Maddie Silcock, Soumyadeep Das, Shravya Shenoy, and Akshara Binu for their helpful discussion regarding this work.

\section*{Data Availability}
The VLT/Xshooter data presented and analysed in this work is available from the ESO Science Archive Facility (\url{https://archive.eso.org/eso/eso_archive_main.html}) with Run/Programme ID 111.24FW.001
0111.B-2322(A), PI Holden. The data used to produce combined VLA+eMERLIN imaging is available from the National Radio Astronomy Observatory VLA (\url{https://science.nrao.edu/facilities/vla/archive/index}) and e-MERLIN (\url{https://www.e-merlin.ac.uk/archive/}) archives with the project IDs given in Section\;2 of \citet{Mutie2024}; reduced and calibrated data is available upon reasonable request to the corresponding author of that study. The HST/WFPC2 F502N imaging shown in Figure\;\ref{fig: slit_position} is available via the Hubble Legacy Archive (\url{https://hla.stsci.edu/}) under proposal ID 5754, PI Ford.



\bibliographystyle{mnras}
\bibliography{precise_outflow_diagnostics} 

\newpage



\appendix

\section{Spatially-resolved $\mathrm{[O\,III]}\lambda\lambda4959,5007$ profile for NGC\;1068}

\begin{figure*}
    \centering
    \includegraphics[width=\textwidth]{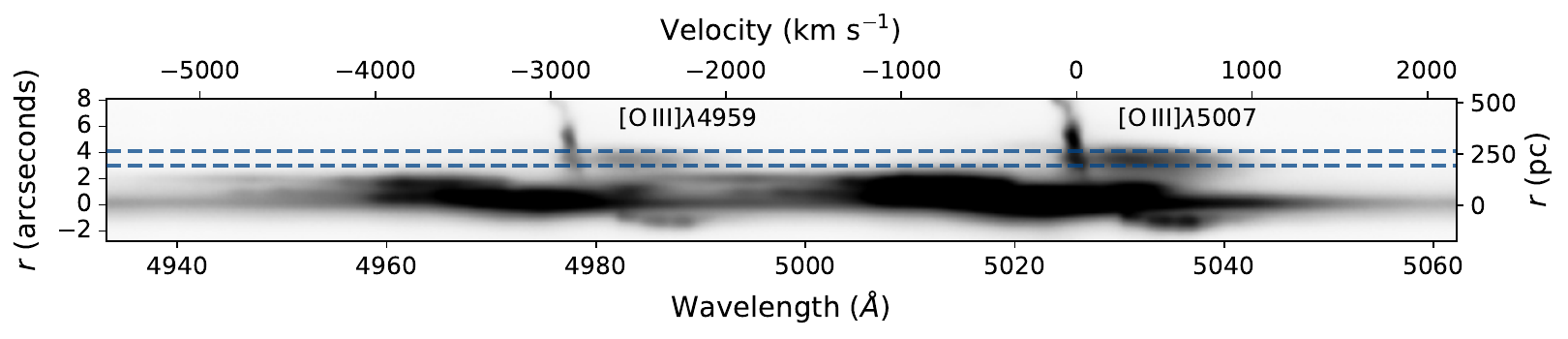}
    \caption{The full kinematic structure of the $\mathrm{[O\,III]}\lambda\lambda4959,5007$ doublet as seen across our Xshooter slit. Here, a radial distance of 0\;arcseconds corresponds to the position of the central AGN; the location of the spectrum that we extract and analyse in the main text is shown by dashed blue lines. Lupton flux scaling \citep{Lupton2004} has been applied here for presentation purposes.}
	\label{fig: oiii_xshooter_full}
\end{figure*}

In Figure\;\ref{fig: oiii_xshooter_full}, we present the entire profile of the $\mathrm{[O\,III]}\lambda\lambda4959,5007$ doublet across the full 11\;arcsecond-extent of the slit (which is centred on the nucleus of NGC\;1068 and aligned along $\mathrm{PA}=38^\circ$: see Figure\;\ref{fig: slit_position}). The location of the extracted spectrum that we analyse in this work is marked with blue dashed lines.

\bsp	
\label{lastpage}
\end{document}